\documentclass[aip,apl,reprint]{revtex4-1}
\usepackage{siunitx}  
\usepackage{subfigure}
\usepackage{graphicx}
\usepackage{amsmath}

\begin{document} 

% Use the \preprint command to place your local institutional report number 
% on the title page in preprint mode.
% Multiple \preprint commands are allowed.
%\preprint{} 
 
\title{Magneto photoluminescence
measurements of tungsten
disulphide monolayers} %Title of paper
 
% repeat the \author .. \affiliation  etc. as needed
% \email, \thanks, \homepage, \altaffiliation all apply to the current author.
% Explanatory text should go in the []'s, 
% actual e-mail address or url should go in the {}'s for \email and \homepage.
% Please use the appropriate macro for the type of information

% \affiliation command applies to all authors since the last \affiliation command. 
% The \affiliation command should follow the other information.

\author{Jan Kuhnert}
%\email[]{Your e-mail address}
%\homepage[]{Your web page}
%\thanks{}
%\altaffiliation{}
\affiliation{Faculty of Physics and Materials Sciences Center,
Philipps-University 35032 Marburg, Germany}

\author{Arash Rahimi-Iman}
\affiliation{Faculty of Physics and Materials Sciences Center,
Philipps-University 35032 Marburg, Germany}
\author{Wolfram Heimbrodt}
\email{Wolfram.Heimbrodt@physik.uni-marburg.de}
\affiliation{Faculty of Physics and Materials Sciences Center,
Philipps-University 35032 Marburg, Germany}

% Collaboration name, if desired (requires use of superscriptaddress option in \documentclass). 
% \noaffiliation is required (may also be used with the \author command).
%\collaboration{}
%\noaffiliation

\date{\today}

\begin{abstract}
Layered transition-metal dichalcogenides have attracted great interest in the
last few years.  Thinned down to the monolayer limit they change from an
indirect band structure to a direct band gap in the visible region. Due to the
monolayer thickness the inversion symmetry of the crystal is broken and spin
and valley are coupled to each other. The degeneracy between the two equivalent
valleys, K and K', respectively, can be lifted by applying an external magnetic
field. Here, we present photoluminescence measurements of CVD-grown tungsten
disulphide (WS${_2}$) monolayers at temperatures of \SI{2}{\kelvin}. By applying
magnetic fields up to \SI{7}{\tesla} in Faraday geometry, a splitting of the
photoluminescence peaks can be observed. The magnetic field dependence of the
A-exciton, the trion and three bound exciton states is discussed and the
corresponding g-factors are determined.
\end{abstract}

%\pacs{}% insert suggested PACS numbers in braces on next line

\maketitle %\maketitle must follow title, authors, abstract and \pacs

% Body of paper goes here. Use proper sectioning commands. 
% References should be done using the \cite, \ref, and \label commands
\section{Introduction}
In the last few years, a new class of monolayer materials has emerged after the
succesful rise of graphene:
transition metal dichalcogenides (TMDCs). The most popular representatives are MoS${_2}$, 
MoSe${_2}$, WS${_2}$, and WSe${_2}$. They have attracted great interest from
physicists all over the world due to their outstanding physical
properties \cite{Mak2010,Sie2014,Ugeda2014,Chernikov2015,Zhao2013,He2014}.
Thinned down to single layers, they change from an indirect band gap
semiconductor in the bulk to a direct band gap semiconductor in the monolayer
regime with its direct band gap located in the K and K' valley of the
Brillouin zone \cite{Splendiani2010,Ellis2011}. The optical properties of these
materials have already been widely studied. One very interesting aspect, to name
but a few, is that in the monolayer limit the spin and valley degrees of freedom
are coupled. This allows to selectively address the K or the K' valley by the helicity of the incident
light \cite{Li2014,MacNeill2015,Smolenski2016,Cao2012}. To break the
degeneracy of the energetically equivalent K and K' valleys, an external
magnetic field can be applied that couples to the magnetic moments perpendicular to the hexagonal
lattice structure of each valley. Due to the different sign of the magnetic
moments in the K and K' valley, the two components split into two Zeeman
components which spectrally can be distinguished. Up to now investigations on
the magnetic properties of transition metal diselenides
have been reported
\cite{Li2014,MacNeill2015,Aivazian2015,Srivastava2014,Mitioglu2015,Mitioglu2016}.\\
So far, the excitonic g-factors for monolayers have been determined for
neutral excitons (A exciton) to be ($-3.94\pm\,0.04$)
in WS${_2}$ \cite{Stier2015}, ($-4.0\pm\,0.2$)
in MoS${_2}$ \cite{Stier2015}, ($-3.7\pm\,0.2$), ($-4.37\pm\,0.15$),
($-4.0\pm\,0.5$), ($-1.57\ \text{upto} -2.86$) in
WSe${_2}$ \cite{Wang2015,Srivastava2014,Mitioglu2015,Aivazian2015}, and
($-3.8\pm\,0.2$) \cite{MacNeill2015,Wang2015} and ($-4.1\pm\,0.2$)
\cite{Li2014} in MoSe${_2}$, respectively.
Reported values for the g-factor of the second excitonic transition (B exciton) are only
available for bulk material: ($-3.99\pm\,0.04$) for WS${_2}$ and
($-4.65\pm\,0.17$) for MoS${_2}$ \cite{Stier2015}. For charged excitons (trions)
in WSe${_2}$ effective g-factors of ($-4.0\pm\,0.5$) and
($-6.28\pm\,0.32$) \cite{Mitioglu2015,Srivastava2014} are reported, and for
MoSe${_2}$  g-factors of ($-3.8\pm\,0.2$) \cite{MacNeill2015} and
($-4.1\pm\,0.2$) upto ($-6.2\pm\,0.2$) depending on the doping
level \cite{Li2014} are reported, respectively. To our knowledge there have been
no reports on trionic g-factors for transition metal disulphides yet.\\
It is worth to note that the photoluminescence (PL) of the TMDCs, however, often
consists of various bands:
Besides the A-exciton and trion, the bi-exciton and even bound exciton states
can be observed, which can show a similar but yet not identical behaviour in
external fields. In this work we perform polarization-resolved
PL measurements of WS${_2}$ monolayers in magnetic fields up to \SI{7}{\tesla},
in order to determine the g-factors of these various excitonic species in this
host lattice.
%%%%%%%%%%%%%%%%%%%%%%%%%%%%%%%%%%%%%%%%%%%%%%%%%%%%%%%%%%%%%%%%%%%%%%%%%%%%%%%%%%%%%%%%%%%%%%%%%%%%%%%%%%%%%%%%%%
%%%%%%%%%%%%%%%%%%%%%%%%%%%%%%%%%%%%%%%%%%%%%%%%%%%%%%%%%%%%%%%%%%%%%%%%%%%%%%%%%%%%%%%%%%%%%%%%%%%%%%%%%%%%%%%%%%
\section{Materials \& methods}
\subsection{Sample}
The WS${_2}$ monolayer sample was grown via CVD on an opaque
silicon substrate covered by SiO${_2}$. The substrate size is approximately $20
\times$\SI{10}{\milli\meter^2} and was purchased from 2Dsemiconductors
Inc. The WS${_2}$ monolayers
are triangularly shaped and have a size of roughly
$30\times$\SI{30}{\micro\meter^2}.
%\footnote{\url{http://www.2dsemiconductors.com/}}
\subsection{Raman measurements}
\begin{figure}[t]
\includegraphics[width=1\columnwidth]{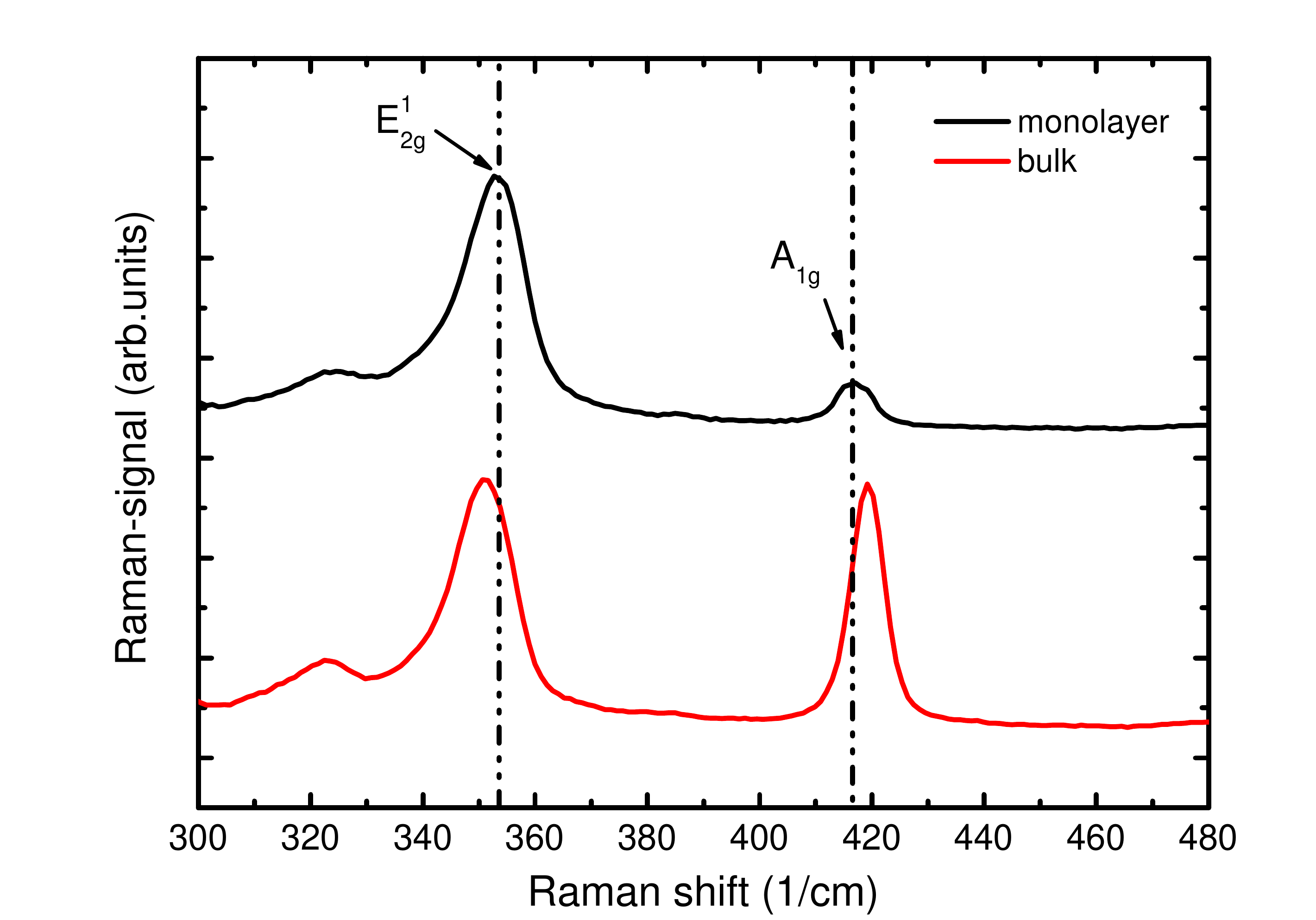}
\caption{Raman measurement of the WS${_2}$ sample with the
characteristic $A_{1g}$ and $E^1_{2g}$ raman lines. The smaller energy
difference for the monolayer mode ($\SI{63,46}{\per\centi\meter}$) can be clearly
seen compared to the bulk mode ($\SI{68}{\per\centi\meter}$). The experiments
were carried out at room temperature.}
\label{fig1} 
\end{figure}
To find the monolayer flakes on the sample raman measurements were performed
prior to measuring the magneto PL. The raman measurements were
carried out at room temperature using a standard Olympus BX41 microscope
equipped with a raman setup. A \SI{80}{\milli\watt} small band \SI{514.5}{\nano\meter} argon-ion
laser was focused on the sample via a 100\,x standard objective . The
holographic notch filter allows to measure raman shifts bigger than
\SI{200}{\centi\meter^{-1}} relative to the laser wavenumber which is
sufficient to measure the characteristic $A_{1g}$ and $E^1_{2g}$ raman lines at
$\sim$ 420 \& \SI{350}{\centi\meter^{-1}}, respectively.
\subsection{Magneto luminescence measurements}
For the magneto-PL (M-PL) measurements, the sample was glued into an
Oxford crysotate that is equipped with a superconducting liquid helium cooled
magnet delivering fields up to 7 Tesla.
 The fields can either be applied in Voigt or in Faraday geometry. For our
 measurements, we solely used the Faraday geometry, for which the
 applied magnetic field is parallel to the k-vektor of the emitted light field
 and therefore perpendicular to the sample surface. The sample was excited by a
 \SI{150}{\milli\watt} \SI{532}{\nano\meter} diode-pumped intracavity-frequency-
 doubled Nd:YAG laser, which was focussed onto the sample via a
 lens to a spot size of approximately $25\times$\SI{25}{\micro\meter^2}.
   PL was collected, collimated and focussed via lenses onto the
    \SI{10}{\micro\meter} entrance slit of a \SI{1250}{\milli\meter}
    spectrometer equipped with a 1200-lines grating and detected by a highly
    sensitive thermo-electrically-cooled silicon CCD.
     In order to distinguish between $\sigma^+$- and $\sigma^-$-polarized light,
     a linear-polarizer plate was placed in front of the entrance slit of the
     spectrometer and a rotateable quarter-wave plate in front of the polarizer
     to select the desired circular polarization. All measurement were carried
     out at a sample temperature of \SI{2}{\kelvin}.
     
\begin{figure}[t]
\includegraphics[width=1\columnwidth]{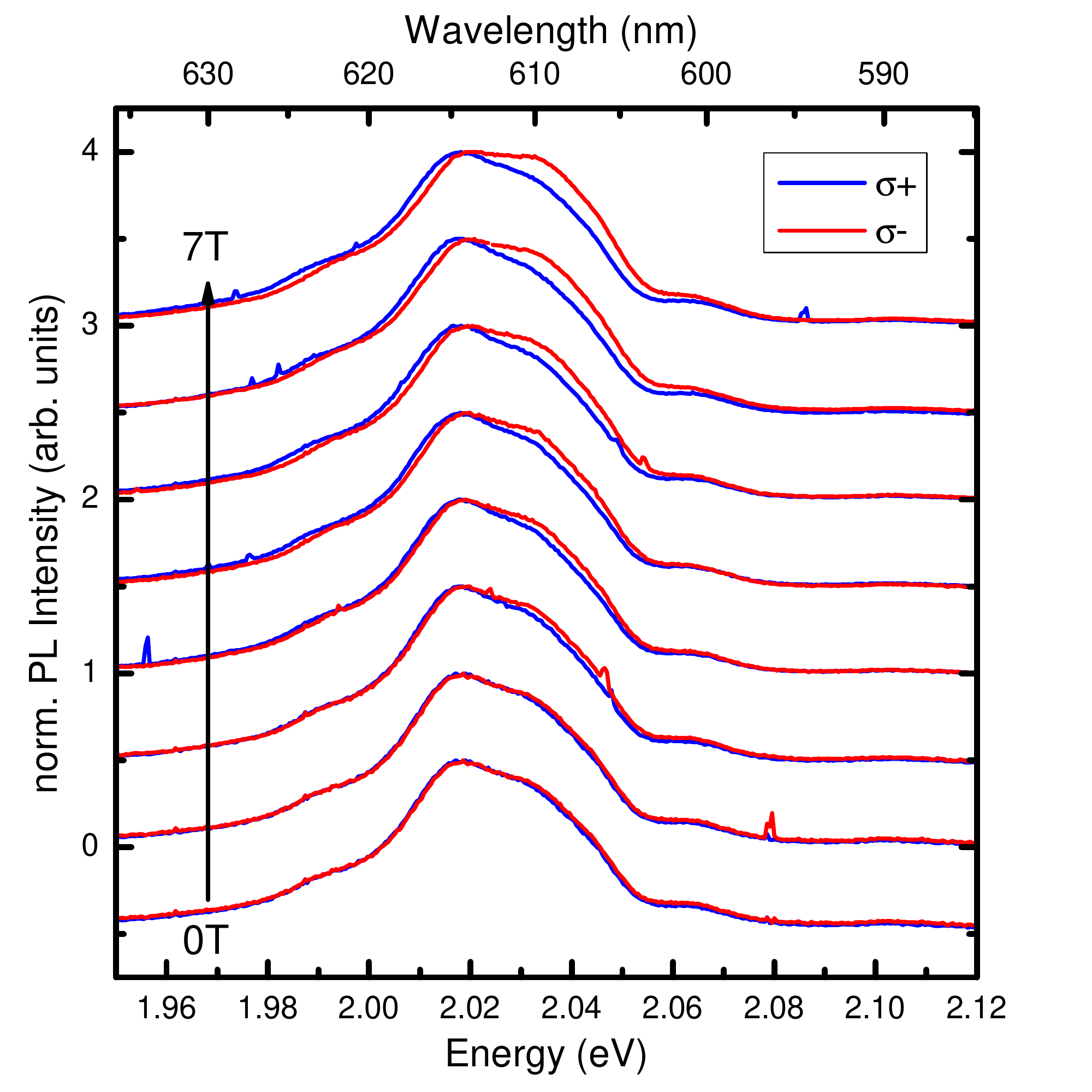}
\caption{Photoluminescence measurements of the WS${_2}$ monolayer with applied
magnetic fields up to 7 Tesla in Faraday geometry using a
\SI{532}{\nano\meter} laser for excitation.
The PL is detected in a $\sigma^+$- and $\sigma^-$-polarized basis. The sample
temperature was kept at \SI{2}{\kelvin}. The spectra are normalized and shifted
vertically for clarity.}
\label{fig2}  
\end{figure}
     
\section{Results \& discussion}
Raman spectroscopy is an excellent tool to identify the number of layers in
few-layer TMDC flakes. Figure \ref{fig1} displays typical Raman-spectra of such samples.
Here, the Raman signal of bulk WS${_2}$ is compared with the one from our
CVD sample, both recorded at room temperature.
The $A_{1g}$ line, which is the out of plane mode, is located at
\SI{416,70}{\per\centi\meter} and the $E^1_{2g}$ line, the inplane mode,
is located at \SI{353,24}{\per\centi\meter} in case of the CVD sample,
respectively.
The difference between both modes (indicated by dashed vertical lines in the
diagram) is \SI{63,46}{\per\centi\meter} which is a good indicator for a
monolayer regime.
The respective separation in the case of the bulk sample is
\SI{68}{\per\centi\meter}. The spectra are normalized to the $E^1_{2g}$ line
for the sake of clarity.
The observed change in the relative separation ratio has been reported
\cite{Lee2010} and has been attributed to dielectric screening
\cite{Molina-Sanchez2011}.\\
The M-PL measurements at a temperature of \SI{2}{\kelvin} are
shown in figure \ref{fig2} for left-circularly ($\sigma^-$, red solid lines) and
right-circularly polarized ($\sigma^+$, blue solid lines) luminescence following 
laser excitation at \SI{2,33}{\electronvolt}. The pump density amounted to
\SI{15}{\kilo\watt\per\centi\meter\squared}. The spectra are normalized and
offset by a constant value to form a waterfall diagram.
The increasing splitting of the PL peaks with increasing magnetic field strength
is clearly visible, here. 

It is also apparent that with increasing fields the $\sigma^-$-component
grows relative to the $\sigma^-$+component, becoming more pronounced on the
higher-energy side of its peak.\\ 
\begin{figure}[t]%
\subfigure{\includegraphics[width=0.49\columnwidth]{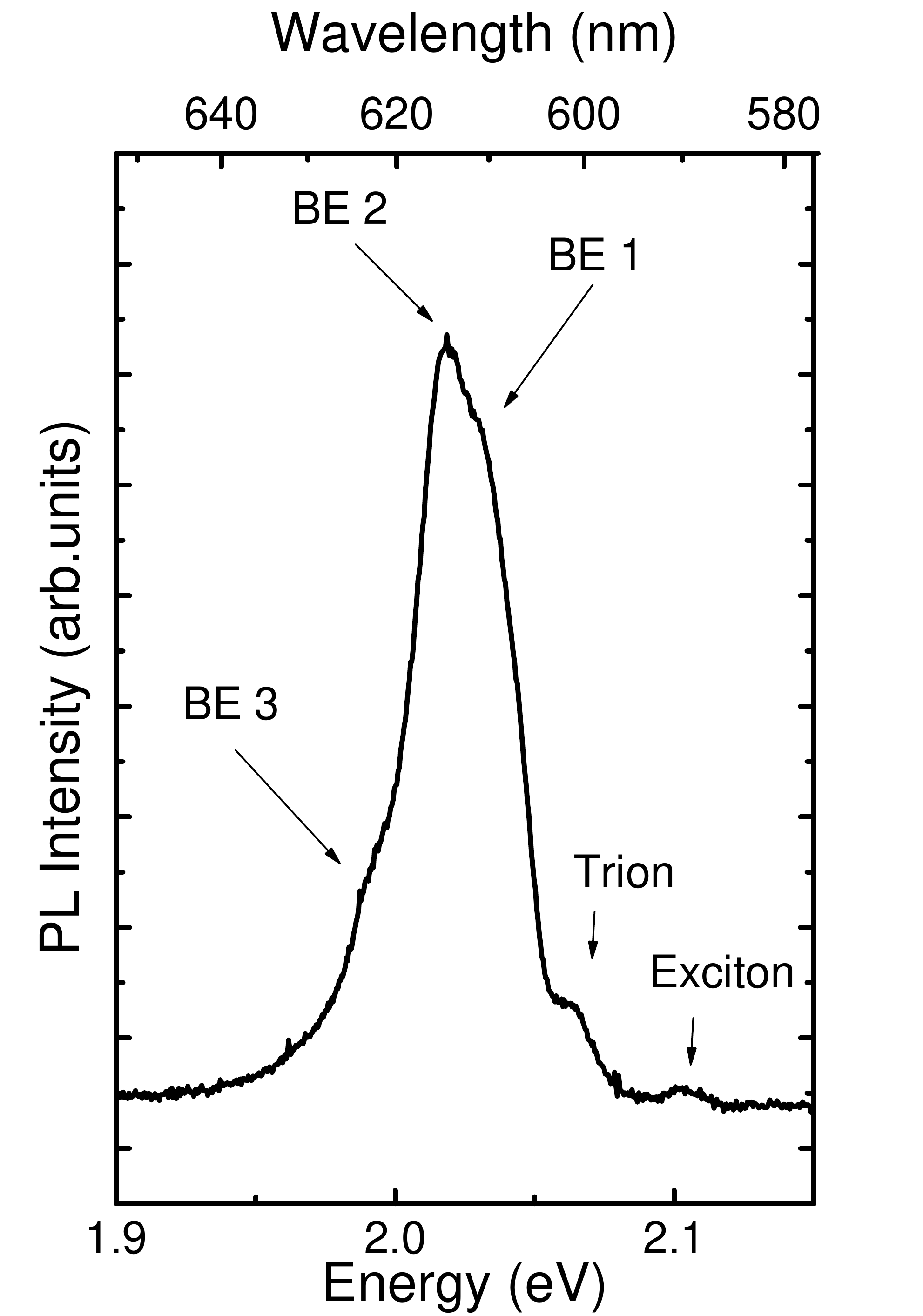}}
\subfigure{\includegraphics[width=0.49\columnwidth]{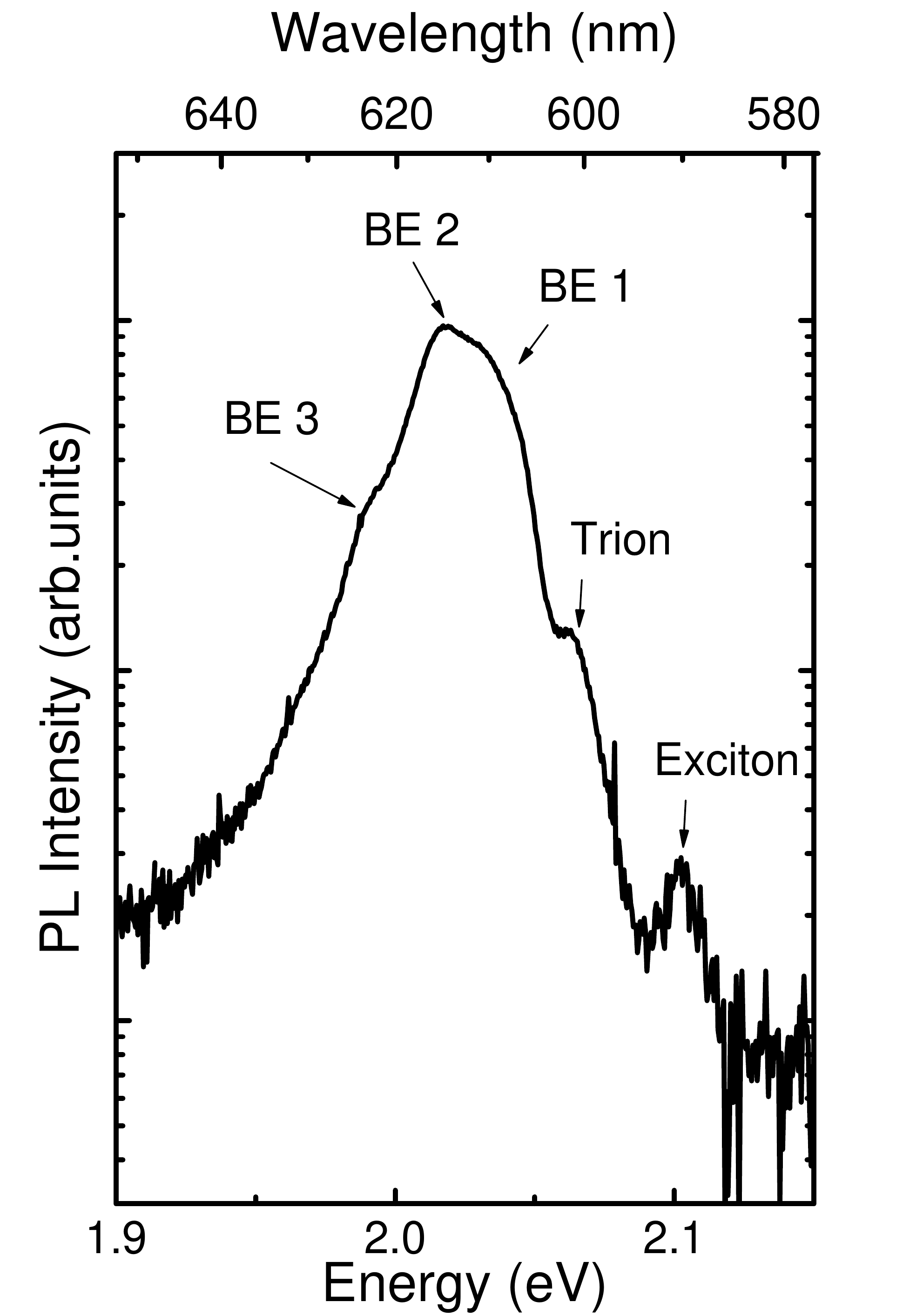}}
\caption{Photoluminescence spectrum of the WS${_2}$ monolayer sample (linear
plot on the left and logarithmic plot on the right hand side) at a temperature
of \SI{2}{\kelvin} without applied magnetic field. The logarithmic plot provides
a more distinct view of the exciton and trion peaks.}
\label{fig3}
\end{figure}
It is obvious that the obtained spectrum is governed by an overlap of various
bands. To get a deeper insight in the individual PL bands our representative spectrum
is plotted both in linear (left) and logarithmic scale (right) in figure
\ref{fig3}. We denote the individual peaks as generally accepted.  
The highest-energetic peak at \SI{2.1}{\electronvolt} is assigned to the
exciton in agreement with earlier papers\cite{Plechinger2015,Chernikov2014},
where the exciton was found at \SI{2.09}{\electronvolt} at a temperature of
\SI{5}{\kelvin}. The trion emission peak is centered at
\SI{2.06}{\electronvolt}.
The visibility of the exciton and trion peaks at elevated energies indicates a
high quality of the CVD sample, as this peak is absent in some earlier studies
\cite{Mitioglu2013,Kioseoglou2015}. 
The most intense PL contribution (centered at \SI{2.025}{\electronvolt}) stems from a superposition
of three bands which we ascribe to bound exciton transitions (labeled BE1, BE2 and BE3).
\begin{figure}[t] 
\includegraphics[width=\columnwidth]{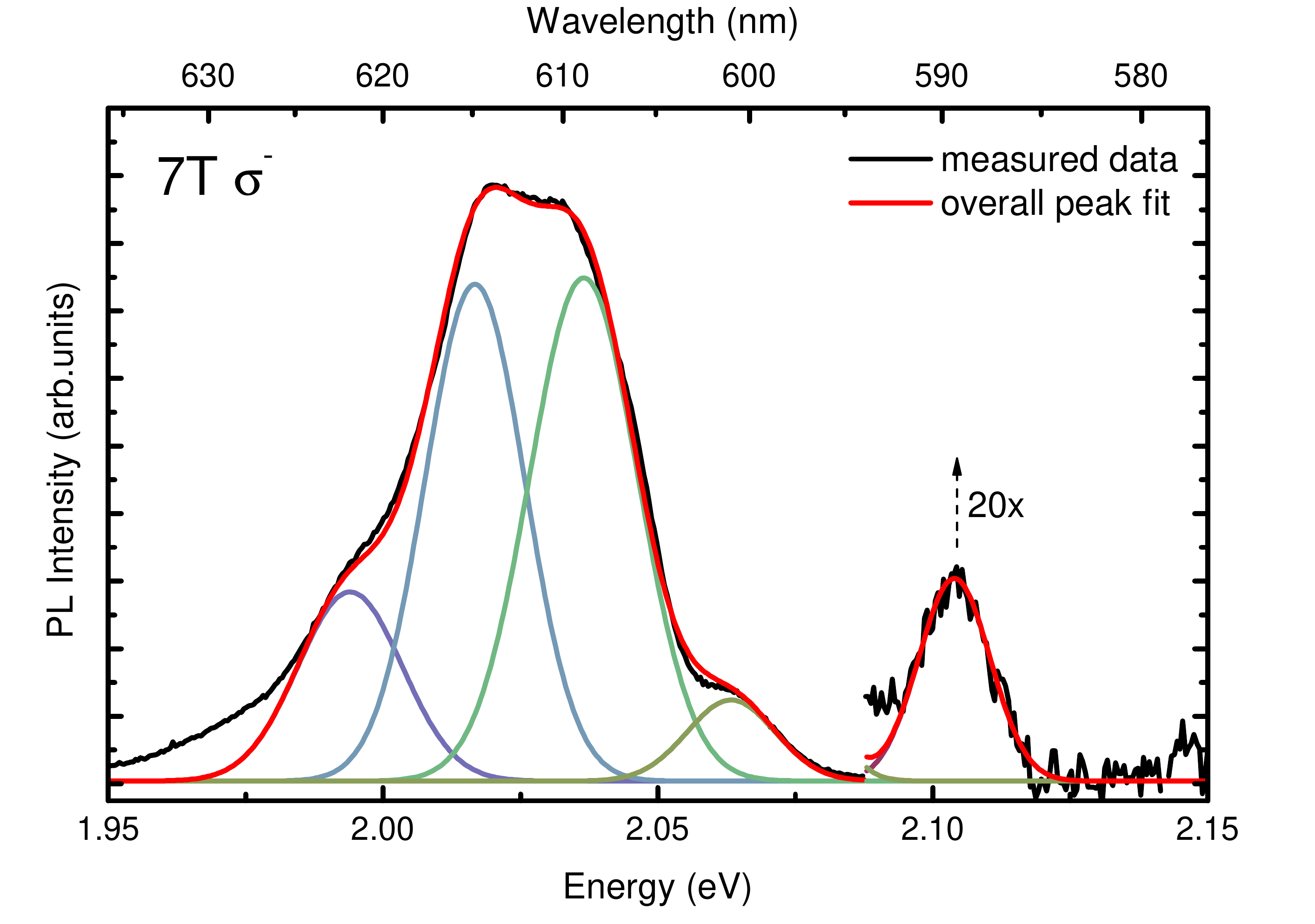}
\includegraphics[width=\columnwidth]{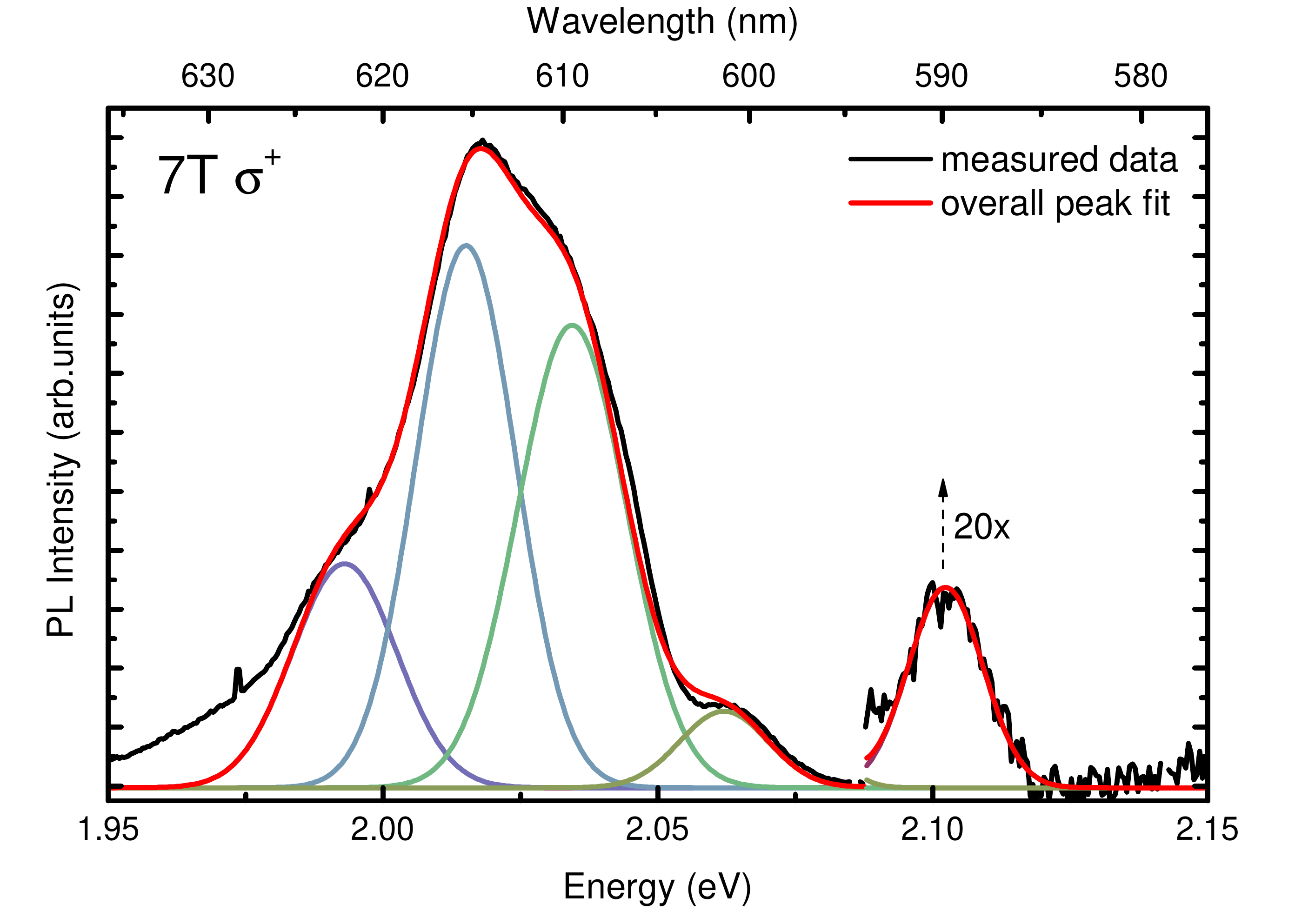}
\caption{Phololuminescence of the WS${_2}$ monolayer with applied magnetic field
of \SI{7}{\tesla} for detection of $\sigma^-$ (top) and $\sigma^+$ (bottom)
polarization of light. The peak components are fitted by using gaussian curves.}
\label{fig4} 
\end{figure} 
By applying a magnetic field in Faraday geometry, the
$\sigma^-$-component shifts towards higher energies, while the
$\sigma^+$-component is shifted to lower energies. Exemplarily, we show the
spectra for the highest applied field of \SI{7}{\tesla} in $\sigma^-$- (top) and
$\sigma^+$ (bottom) circular polarization in figure \ref{fig4}. To determine the
effective g-factors of all the peaks, the spectra have been fitted by means of
multiple gaussian shaped curves.
One gaussian curve is used for each peak component, i.e. exciton, trion,
 defect-bound exciton 1,2,3, respectively. \\
Firstly, the PL spectrum in the case of zero external magnetic field was fitted
with five gaussians to determine both the energetic positions and the full width
at half maximum (FWHM) values.
The exciton and trion peaks exhibits a substantially smaller FWHM
(\SI{16.95}{\milli\electronvolt} and \SI{18.84}{\milli\electronvolt}) than the
defect-bound exciton-emissions (\SI{23.19}{\milli\electronvolt},
\SI{22.37}{\milli\electronvolt}, and \SI{21.38}{\milli\electronvolt}).
The FWHMs were kept unchanged for the fitting procedure in higher fields to reduce the number of free parameters.
By fitting the $\sigma^-$ and $\sigma^+$ \SI{7}{\tesla} spectra, we could
determine the peak positions and amplitudes of the five bands. The best fitting
results for both 7T $\sigma^-$(top) and 7T $\sigma^+$(bottom) are shown in
figure \ref{fig4}. 
It should be mentioned that the diamagnetic shift of the transitions
does not need to be taken into account in a field range up to \SI{7}{\tesla}
for these materials. Higher fields in the region of $50$ to \SI{100}{\tesla} are needed for
remarkable effects \cite{Stier2015}.\\
The Zeeman splitting is given by the absolute difference between $\sigma^-$ and
$\sigma^+$ of the center positions of the gaussian peaks for each individual
species. Accordingly, for the excitonic transition we found a splitting per Tesla of
\SI{22.86}{\micro\electronvolt\per\tesla}. The corresponding g-factor for the
excitonic mode can be obtained by the equation:
\begin{equation*}
g=\frac{2(E_+-E_-)}{\mu_B B},
\end{equation*} 
Here, $g_{\text{exc}}=-3.95$ is in good agreement with previously reported
values for WS${_2}$ \cite{Stier2015}. For the trion peak, we got a total splitting of 
\SI{23}{\micro\electronvolt\per\tesla} with the resulting
g-factor of $g_{\text{trion}}=-3.97$ very close to the free exciton value. An
identical value for excitons and trions has been reported also for the
selenides \cite{Li2014,MacNeill2015,Srivastava2014,Mitioglu2015}.
Basically a very similar or identical value can be expected for excitons and
trions, since the splitting arises mainly due to the difference in the orbital
magnetic moment of the initial and the final state. The spin contribution is
expected to be zero, since the optical transitions take place between bands having the same spin.
It should be mentioned, that there are also reports about a slightly higher
g-value\cite{Srivastava2014,Aivazian2015} for the trion.\\
It is interesting to note, that the three bound exciton states exhibit different
g-values. We find a splitting of \SI{25,29}{\micro\electronvolt\per\tesla} for
the BE1 centered at \SI{2.035}{\electronvolt} which gives a g-factor of
$g_{BE1}=-4,37$. This value is slightly higher than the values for the free exciton transition.
The second bound exciton-emission peak (BE2) centered at
\SI{2.0159}{\electronvolt} shows a shift of
\SI{23.14}{\micro\electronvolt\per\tesla} yielding a g-factors of
$g_{BE2}=-4,00$ in agreement with the free exciton value but the third band
(BE3) centered at \SI{1.9935}{\electronvolt} exhibits
\SI{14.28}{\micro\electronvolt\per\tesla} with $g_{BE3}=-2.47$.
The accuracy for the estimation of the g-factors is $\pm\,0.4$ in our analysis,
mainly caused by the uncertainty of the gaussian peak determination.
A different splitting for different sample positions, as mentioned in
Ref\cite{Aivazian2015}, could not be observed in our CVD grown sample.
Our results strongly indicate a tendency: lower emission energies of the bound
excitons are due to larger binding energies to the defect, which are most likely
neutral or ionised donor or acceptor states.
Larger binding energies and respective stronger localization of the excitons
yield obviously smaller g-values. We ascribe these changes to different
valley orbital contributions to the magnetic moments of the conduction and
valence band. The simple two-band $k \cdot p$ approximation with $m_e=m_h$ does
not affect the transition energies. Corrections beyond the two-band model give
different effective masses and respectively  different valley magnetic moments
for the electrons and holes \cite{Liu2013,Kormanyos2013}.
   %%%%%%%%%%%%%%%%%%%%%%%%%%%%%%%%%%%%%%%%%%%%%%%%%%%%%%%%%%%%%%%%%%%%%%%%%%%%%%%%%%%%%%%%%%%%%%%%%%%%%%%%%%%%%%%%%%
%%%%%%%%%%%%%%%%%%%%%%%%%%%%%%%%%%%%%%%%%%%%%%%%%%%%%%%%%%%%%%%%%%%%%%%%%%%%%%%%%%%%%%%%%%%%%%%%%%%%%%%%%%%%%%%%%%
\section{Conclusion} 
To summarize, we investigated the magneto-PL of CVD grown WS${_2}$ monolayers in
magnetic fields up to 7 Tesla.
In Faraday geometry, we revealed the Zeeman-splitting in polarization-sensitive
spectra for five excitonic modes, the A-exciton, the trion and three bound exciton states. 
We determined the A-exciton g-factor to be $g_{exc}=-3.95\pm\,0.4$, the trion g-factor to be
$g_{trion}=-3.97\pm\,0.4$.  
For the bound excitons we could determine deviations from the free exciton
value, which should be caused by different effective masses of the electrons and
holes due to different localization lengths.
%%%%%%%%%%%%%%%%%%%%%%%%%%%%%%%%%%%%%%%%%%%%%%%%%%%%%%%%%%%%%%%%%%%%%%%%%%%%%%%%%%%%%%%%%%%%%%%%%%%%%%%%%%%%%%%%%%
%%%%%%%%%%%%%%%%%%%%%%%%%%%%%%%%%%%%%%%%%%%%%%%%%%%%%%%%%%%%%%%%%%%%%%%%%%%%%%%%%%%%%%%%%%%%%%%%%%%%%%%%%%%%%%%%%%
% If in two-column mode, this environment will change to single-column format so that long equations can be displayed. 
% Use only when necessary.
%\begin{widetext}
%$$\mbox{put long equation here}$$
%\end{widetext}

% Figures should be put into the text as floats. 
% Use the graphics or graphicx packages (distributed with LaTeX2e).
% See the LaTeX Graphics Companion by Michel Goosens, Sebastian Rahtz, and Frank Mittelbach for examples. 
%
% Here is an example of the general form of a figure:
% Fill in the caption in the braces of the \caption{} command. 
% Put the label that you will use with \ref{} command in the braces of the \label{} command.
%
% \begin{figure}
% \includegraphics{}%
% \caption{\label{}}%
% \end{figure}

% Tables may be be put in the text as floats.
% Here is an example of the general form of a table:
% Fill in the caption in the braces of the \caption{} command. Put the label
% that you will use with \ref{} command in the braces of the \label{} command.
% Insert the column specifiers (l, r, c, d, etc.) in the empty braces of the
% \begin{tabular}{} command.
%
% \begin{table}
% \caption{\label{} }
% \begin{tabular}{}
% \end{tabular}
% \end{table}

% If you have acknowledgments, this puts in the proper section head.
\begin{acknowledgments}We gratefully acknowledge financial support of the German Science Foundation
(DFG) in the framework of the SFB 1083. Furthermore, supply of high-quality
samples for this study by 2Dsemiconductors Inc. is acknowledged.
\end{acknowledgments}
 
% Create the reference section using BibTeX:
\bibliography{references_magneto_PL_WS2}
     
\end{document}